\documentstyle[prd,aps,floats]{revtex}

\def\be{\begin{equation}}
\def\ee{\end{equation}}
\def\bea{\begin{eqnarray}}
\def\eea{\end{eqnarray}}

\begin{document}
\preprint{SUSX-TH-97-022, SUSSEX-AST 97/11-1, PU-RCG/97-20, gr-qc/9711068}
\draft

\input epsf
\renewcommand{\topfraction}{0.99}
\renewcommand{\bottomfraction}{0.99}
\twocolumn[\hsize\textwidth\columnwidth\hsize\csname
@twocolumnfalse\endcsname

\title{Quintessence models of Dark Energy with non-minimal coupling}
\author{Tame Gonzalez, Genly Leon and Israel Quiros}
\address{Central University of Las Villas. Santa Clara. Cuba}

\date{\today}
\maketitle
\begin{abstract}
We explore quintessence models of dark energy which exhibit
non-minimal coupling between the dark matter and the dark energy
components of the cosmic fluid. The kind of coupling chosen is
inspired in scalar-tensor theories of gravity. We impose a
suitable dynamics of the expansion allowing to derive exact
Friedmann-Robertson-Walker solutions once the coupling function is
given as input. Coupling functions that lead to self-interaction
potentials of the single and double exponential types are the
target of the present investigation. The stability and existence
of the solutions is discussed in some detail. The models are
tested against the observational evidence.
\end{abstract}

\pacs{PACS numbers: 04.20.Jb, 04.20.Dw, 98.80.-k, 98.80.Es,
95.30.Sf, 95.35.+d}

\vskip2pc]

\section{Introduction}

At the time, all the available observational evidence suggest that
the energy density of the universe today is dominated by a
component with negative pressure of, almost, the same absolute
value as its energy density. This mysterious component, that
violates the strong energy condition (SEC), drives a present
accelerated stage of the cosmic evolution and is called "dark
energy" (DE). A pleiad of models have been investigated to account
for this SEC violating source of gravity. Among them, one of the
most successful models is a slowly varying scalar field, called
'quintessence'\cite{kolda,quintessence}.

Many models of quintessence assume that the background and the
dark energy evolve independently, so their natural generalization
are models with non-minimal coupling between both components.
Although experimental tests in the solar system impose severe
restrictions on the possibility of non-minimal coupling between
the DE and ordinary matter fluids \cite{will}, due to the unknown
nature of the dark matter (DM) as part of the background, it is
possible to have additional (non gravitational) interactions
between the DE and the DM components, without conflict with the
experimental data.\footnotemark\footnotetext{It should be pointed
out, however, that when the stability of dark energy potentials in
quintessence models is considered, the coupling dark matter-dark
energy is troublesome \cite{doram}.} In the present paper we will
be concerned, precisely, with non-minimal coupling between the DM
and DE components of the cosmological fluid. Since, models with
non-minimal coupling imply interaction (exchange of energy) among
the DM and the DE, these models provide new qualitative features
for the coincidence problem \cite{luca,pavon}. It has been shown,
in particular, that a suitable coupling, can produce scaling
solutions. The way in which the coupling is approached is not
unique. In reference \cite{luca}, for instance, the coupling is
introduced by hand. In \cite{pavon} the type of coupling is not
specified from the beginning. Instead, the form of the interaction
term is fixed by the requirement that the ratio of the energy
densities of DM and quintessence be a constant. In
\cite{chimento}, a suitable interaction between the quintessence
field and DM leads to a transition from the domination matter era
to an accelerated expansion epoch in the model of
Ref.\cite{pavon}. The authors obtained new features for the
coincidence problem. A model derived from the Dilaton is studied
in \cite{bean1}. In this model the coupling function is chosen as
a Fourier expansion around some minimum of the scalar field.

A variety of self-interaction potentials  have been studied. Among
them, a single exponential is the simplest case. This type of
potential leads to two possible late-time attractors in the
presence of a barotropic fluid\cite{wands,copeland}: i) a scaling
regime where the scalar field mimics the dynamics of the
background fluid, i. e., the ratio between both DM and
quintessence energy densities is a constant and ii) an attractor
solution dominated by the scalar field. Given that single
exponential potentials can lead to one of the above scaling
solutions, then it should follow that a combination of
exponentials should allow for a scenario where the universe can
evolve through a radiation-matter regime (attractor i)) and, at
some recent epoch, evolve into the scalar field dominated regime
(attractor ii)). Models with single and double exponential
potentials has been studied also in references \cite{ruba,Isra1}.

The aim of the present paper is to investigate models with
non-minimal coupling among the components of the cosmic fluid, in
which, single and double exponential self interaction potentials
are involved. To specify the coupling, we take as a Lagrangian
model, a scalar-tensor theory with Lagrangian written in Einstein
frame variables. In this paper, additionally, we impose the
dynamics of the expansion by exploring a linear relationship
between the Hubble parameter and the time derivative of the scalar
field. Using this relationship we can solve the field equations
explicitly. A flat Friedmann-Robertson-Walker (FRW) universe
filled with a mixture of two interacting fluids: DM and
quintessence, is considered.

The paper has been organized as follow. In section 2 the details
of the model are given. The method used to derive FRW solutions is
explained in section 3. We use the approach of 'two fluids': a
background fluid of DM and a self-interacting scalar field
(quintessence). Two different couplings that lead to
self-interaction potentials of single exponential and double
exponential class are studied separately. In section 4 a study of
existence and stability of the solutions is presented. The models
studied in the former sections are tested against the
observational evidence in section 5. Finally, in section 6,
conclusions are given.

\section{The model}

We consider the following action inspired in a scalar-tensor
theory written in Einstein frame, where the quintessence (scalar)
field is coupled to the matter degrees of freedom:

\begin{eqnarray}
S=\int_{M_4}d^4x\sqrt{|g|}\{\frac{1}{2}R-\frac{1}{2}(\nabla\phi)^2\nonumber\\
-V(\phi)+ C^2(\phi){\cal L}_{(matter)}\}.\label{action}
\end{eqnarray}
In this equation $R$ is the curvature scalar, $C^2(\phi)$ - the
coupling function, and ${\cal L}_{matter}$ is the Lagrangian
density for the matter degrees of freedom.

We are concerned here with a flat FRW universe that is filled with
a background pressureless dark matter and a quintessence field
(the scalar field $\phi$). The line element is:

\begin{equation}
ds^2=-dt^2+a(t)^2\delta_{ij}dx^idx^j,
\end{equation}
where $i,j=1,2,3$ and we use the system of units in which $8\pi
G=c=\hbar=1$. The field equations that are derived from the action
(\ref{action}) are:

\begin{eqnarray}
3H^{2}&=&\rho_{m}+\frac{1}{2}\dot\phi^{2}+V\label{frieman1},\\
2\dot H+3H^{2}&=&(1-\gamma)\rho_{m}-
\frac{1}{2}\dot\phi^{2}+V \label{frieman2},\\
\ddot\phi+3H\dot\phi &=&-V'-(\ln{X})'\rho_{m} \label{frieman3},
\end{eqnarray}
where we have introduced the reduced notation $X(\phi)\equiv
C(\phi)^{(3\gamma-4)/2}$. The "continuity" equations for
background (DM) and the quintessence field are:
\begin{eqnarray} \dot\rho_{m}+3\gamma H
\rho_{m}=-(\ln{X})'\dot\phi\;\rho_{m} \label{ecuaciondinamica1},
\end{eqnarray} or, after integration
\begin{eqnarray}
\rho_{m}=Ma^{-3\gamma}X^{-1}\label{ecuaciondinamica2}.\end{eqnarray}
where M is a constant of integration. In the former equations the
dot accounts for derivative in respect to the comoving time $t$,
while the comma denotes derivative in respect to $\phi$.

\section{Deriving Solutions}

Summing up equations (\ref{frieman1}) and (\ref{frieman2}) one
obtains:
\begin{eqnarray} \dot H+3H^{2}=\frac{2-\gamma}{2}\rho_{m}+V.
\label{sumafrieman}\end{eqnarray} Let us fix the dynamics of the
cosmic evolution by introducing the following constrain
\begin{eqnarray} \dot\phi=\lambda H,
\label{restriccion1}\end{eqnarray}which implies
\begin{eqnarray} a=e^{\phi/\lambda},\label{restriccion}\end{eqnarray}
and rewriting of  (\ref{ecuaciondinamica2}) in the form:
\begin{eqnarray}
\rho_{m}(\phi)=Me^{-\frac{3\gamma}{\lambda}\phi}X^{-1}(\phi).
\label{expresionro}\end{eqnarray}

If one substitutes Eqn.(\ref{restriccion1}) in (\ref{frieman3})
and the resulting equation is compared with (\ref{sumafrieman})
then, one obtains a differential equation on the potential $V$,
that relates $V$ and the coupling function $X$ (and their
derivatives):
\begin{eqnarray} \frac{dV}{d\phi}+\lambda
V(\phi)=\rho_{m}(\phi)\left(\frac{d\ln{X}}
{d\phi}-\frac{\lambda(\gamma-2)}{2}\right).\label{equacionvx}\end{eqnarray}

If one realizes that $d\phi=\lambda d(\ln a)$ (see equation
(\ref{restriccion})), then the equation (\ref{equacionvx}) can be
rewritten as:

\begin{eqnarray}
V'+\lambda^{2}V=\left(\frac{X'}{X}-\frac{\lambda(\gamma-2)}{2}
\right)\frac{M}{a^{3\gamma}}X^{-1},\label{equacionvx2}
\end{eqnarray}
where now the comma denotes derivative in respect to the variable
$N=\ln a$. With the constrain (\ref{restriccion1}) and the
equation (\ref{expresionro})  we can integrate the equation
(\ref{frieman1}) in quadratures:

\begin{eqnarray}
\int \frac{d \phi}{\sqrt{M
e^{-\frac{3\gamma}{\lambda}\phi}X^{-1}(\phi)+V(\phi)}}=\sqrt{\frac{2
\lambda^{2}}{6-\lambda^{2}}}(t+t_{0}),\label{exp14}
\end{eqnarray}
or, if one introduces the time variable $d\tau
=e^{-\frac{3\gamma}{2\lambda}\phi}X^{-1/2}dt$, (\ref{exp14}) can
be written as:

\begin{eqnarray}
\int \frac{d\phi}{\sqrt{M
+e^{\frac{3\gamma}{\lambda}\phi}X(\phi)V(\phi)}}= \sqrt{\frac{2
\lambda^{2}}{6-\lambda^{2}}} (\tau+\tau_{0}). \label{exp16}
\end{eqnarray}

In consequence, once the function $X(\phi)$ (or $X(a)$) is given
as input, then one can solve equation (\ref{equacionvx}) (or
(\ref{equacionvx2})) to find the functional form of the potential
$V(\phi)$ (or $V(a)$). The integral (\ref{exp14}) (or
(\ref{exp16})) can then be taken explicitly to obtain $t=t(\phi)$
(or $\tau=\tau(\phi)$). By inversion we can obtain $\phi=\phi(t)$
(or $\phi=\phi(\tau)$) so, the scale factor can be given as
function of either the cosmic time $t$ or the time variable $\tau$
through Eqn.(\ref{restriccion}).

\subsection{Particular Cases}

In this section we study separately two different choices of the
coupling function $X(\phi)$ that lead to double and single
exponential potentials respectively. The importance of this class
of potentials in cosmology has been already outlined in the
introductory part of this paper.

\subsubsection{Case A}

Let us consider the simplest case when
\begin{eqnarray}
\frac{d}{d\phi}\left(\ln{X(\phi)}\right)=const=\alpha\Rightarrow
X(\phi)=X_{0}e^{\alpha\phi},\label{acoplamientose}\end{eqnarray}
where $X_{0}$ is an integration constant. In this case, by
integration of (\ref{equacionvx}) one obtains
\begin{eqnarray}
V(\phi)=V_{0}e^{-\lambda\phi}+W_{0}e^{-(\alpha+3\gamma/\lambda)\phi},
\label{potencialdoble}\end{eqnarray}
where the constant
\begin{eqnarray}
W_{0}=-\frac{2M}{X_{0}}\left(\frac{2\alpha-\lambda
(2-\gamma)}{\alpha+3\gamma/\lambda-\lambda}\right). \label{exp18}
\end{eqnarray}
We are faced with a self-interaction potential that is a
combination of exponentials with different constants in the
exponent. The usefulness of this kind of potential has been
already explained in the introduction and will be briefly
discussed later within the frame of the present model, when we
study the stability of the corresponding solution. If one
introduce the variable $z=e^{-\frac{l}{2}\phi}$, the equation
(\ref{exp16}) can be written as

\begin{eqnarray}
\int \frac{dz}{\sqrt{z^{2}+a^{2}}}= -\frac{l}{2}\sqrt{\frac{2
\lambda^{2}}{6-\lambda^{2}}\frac{X_{0}V_{0}}{a^{2}}}
(\tau+\tau_{0}) \label{exp20}
\end{eqnarray}
where $a^{2}=(X_{0}V_{0})/(M+W_{0}X_{0})$ and
$l=\alpha+3\gamma/\lambda-\lambda$. Integration of (\ref{exp20})
yields to:

\begin{eqnarray}
\phi(\tau)=\phi_{0}+ln\left(\sinh\left[\mu(\tau+\tau_{0})\right]\right)^{-2/l},
\label{exp22}
\end{eqnarray}
and, consequently:
\begin{eqnarray}
a(\tau)=a_{0}\left(\sinh\left[\mu(\tau+\tau_{0})\right]\right)^{-2/l}.
\label{exp23}
\end{eqnarray}
where $\mu=-\frac{l}{2}\sqrt{\frac{2\lambda^{2}}{6-\lambda^{2}}
\frac{X_{0}V_{0}}{a^{2}}}$. On the other hand, by using the
Friedmann equation (\ref{frieman1}) and inserting therein the
potential (\ref{potencialdoble}), one obtains the Hubble parameter
as function of the quintessence field $\phi$:

\begin{eqnarray}
H^{2}(\phi)=\frac{2 V_{0}}{6-\lambda^{2}} \left[ b^{2}
e^{-(\alpha+3 \gamma / \lambda) \phi} +e^{-\lambda \phi} \right],
\label{exp30A}
\end{eqnarray}
where we have considered Eqn.(\ref{expresionro}). In terms of the
scale factor one has, instead,
\begin{eqnarray}
H^{2}(a)=\frac{2 V_{0}}{6-\lambda^{2}} \left[ b^{2} a^{-(\alpha
\lambda+3 \gamma)} +a^{-\lambda^{2}} \right].\label{exp30}
\end{eqnarray}
The former magnitudes can be given in terms of the redshift
variable if one considers that $a(z)=a_{0}/(1+z)$, where
$a_0\equiv a(z=0)$ (for simplicity of the calculations we choose
the normalization $a_{0}=1$):

\begin{eqnarray}
H^{2}(z)=\frac{2 V_{0}}{6-\lambda^{2}} \left[ b^{2}
(z+1)^{-(\alpha \lambda+3 \gamma )} +a^{-\lambda^{2}}.\right]
\label{exp31}
\end{eqnarray}
In redshift variable the energy density of DM is
$\rho_{m}(z)=(M/X_{0})(z+1)^{\alpha \lambda +3 \gamma}=
\rho_{0}(z+1)^{\alpha \lambda +3 \gamma}$, so the DM dimensionless
density parameter can be given as function of $z$ also:
\begin{eqnarray}
\Omega_{m}(z)=\frac{(6-\lambda^{2}) \rho_{0}}{2 V_{0} b^{2}}
\frac{(z+1)^{\alpha \lambda +3 \gamma -
\lambda^{2}}}{(z+1)^{\alpha \lambda +3 \gamma -
\lambda^{2}}+1/b^{2}}.\label{exp33}
\end{eqnarray}

In order to constrain the parameter space of the solution one
should consider the model to fit the observational evidence about
a universe with an early matter dominated period and a former
transition to dark energy dominance. Therefore $\lambda^{2}-\alpha
\lambda -3 \gamma<0$, besides, one can write
$\Omega_{m}(\infty)=(1-\epsilon)$, where $\epsilon$ is a small
number (the small fraction of dark energy component during
nucleosynthesis epoch). In correspondence
$b^{2}=\frac{(6-\lambda^{2})\rho_{0}}{2(1-\epsilon)V_{0}}$. Taking
into account the observational fact that, at present, ($z=0$);
$\Omega_{m}(0)=1/3$, then $1/b^{2}=2-3\epsilon$. After this
Eqn.(\ref{exp33}) can be rewritten in the following way:

\begin{eqnarray}
\Omega_{m}(z)=(1-\epsilon) \frac{(z+1)^{\alpha \lambda +3 \gamma -
\lambda^{2}}}{(z+1)^{\alpha \lambda +3 \gamma - \lambda^{2}}+ (2-3
\epsilon)} \label{exp33A}
\end{eqnarray}

Other physical magnitudes of observational interest are the
quintessence equation of state (EOS) parameter and the
deceleration parameter

\begin{eqnarray}
\omega_{\phi}=-1+\frac{\lambda^{2}/3}{1-\Omega_{m}},
\label{expestado}
\end{eqnarray}
\begin{eqnarray}
q=-1+\frac{\lambda^{2}}{2}+\frac{3 \gamma}{2}\Omega_{m},
\label{expedesa}
\end{eqnarray}
respectively.

\subsubsection{Case B}

A second very simple choice is to consider an exponential coupling
function of the form:
\begin{eqnarray}
X=X_{0} e^{\frac{\lambda (2-\gamma)}{2}\phi}\label{xenB}.
\end{eqnarray}
In this case equation (\ref{equacionvx}) simplifies to

\begin{eqnarray}
\frac{d V}{d \phi} + \lambda V =0, \label{exp34A}
\end{eqnarray}
which can be easily integrated to yield to a single exponential
potential:
\begin{eqnarray}
V=V_{0} e^{-\lambda \phi}. \label{potenB}
\end{eqnarray}
In consequence the equation (\ref{exp16}) can be written as

\begin{eqnarray}
\int \frac{d w}{\sqrt{w^{2}+A^{2}}} = \mu
(\tau+\tau_{0})\label{expr16enB}
\end{eqnarray}
where $w=\exp(-\frac{\gamma (6-\lambda^{2})}{4 \lambda}\phi)$,
$A^{2}=V_{0}X_{0}/M$ and $\mu=\gamma \sqrt{M (\lambda^{2}-6)/8}$,
so we have the explicit solution

\begin{eqnarray}
\phi(\tau)= \phi_{0} + \ln \left( \sinh \left[ \mu
(\tau+\tau_{0})\right]^{\frac{4\lambda}{\gamma (6-\lambda^{2})}}
\right), \label{exprfienB}
\end{eqnarray}
and, consequently:
\begin{eqnarray}
a(\tau)= a_{0} \sinh \left[ \mu
(\tau+\tau_{0})\right]^{\frac{4\lambda}{\gamma (6-\lambda^{2})}}.
\label{expraenB}
\end{eqnarray}

The dimensionless density parameter and the Hubble expansion
parameter are given, as functions of the redshift, through:

\begin{eqnarray}
\Omega_{m}(z)=\frac{6-\lambda^{2}}{6 A^{2}} \frac{(1+z)^{3
\gamma+\gamma (6-\lambda^{2})/2 - \lambda^{2}}}{(1+z)^{3
\gamma+\gamma (6-\lambda^{2})/2 - \lambda^{2}}+A^{2}},
\label{omegaB}
\end{eqnarray}
and
\begin{eqnarray}
H(z)=B\sqrt{\frac{1}{A^{2}}(1+z)^{3 \gamma+\gamma
(6-\lambda^{2})/2}+(1+z)^{\lambda^{2}}},\label{hubbleB}
\end{eqnarray}
respectively, where $B=\sqrt{2 V_{0}/(6-\lambda^{2})}$. Note that,
$\Omega_{m}$ is a maximum ($\Omega_m^\infty=(6-\lambda^2)/6A^2$)
at $z=\infty$. Taking into account that observations are not exact
we admit that, at the epoch of nucleosynthesis
$\Omega_{m}(\infty)=(1-\epsilon)$, where, as before, $\epsilon$ is
a very small number ($\epsilon=[6(A^2-1)+\lambda^2]/6A^2$). Then,
if we assume (as before) that, at present, $\Omega_m(0)=1/3$, then
(\ref{omegaB}) can be rewritten as

\begin{eqnarray}
\Omega_{m}(z)=\frac{(1-\epsilon)(1+z)^{3 \gamma+\gamma
(6-\lambda^{2})/2 - \lambda^{2}}}{(1+z)^{3 \gamma+\gamma
(6-\lambda^{2})/2 - \lambda^{2}}+(2-3\epsilon)} \label{omegaBB}
\end{eqnarray}

Other physical magnitudes of observational interest are the
quintessence EOS parameter and the deceleration parameter, that
are given in equations (\ref{expestado}) and (\ref{expedesa}),
respectively.

\section{Existence and stability of the solutions}

In this section we do not aim at a complete study of the stability
in the case of the double exponential potential. Instead, we use
the following simplifying fact. Note that, in terms of the scale
factor, the potential (\ref{potencialdoble}) can be written as
follows:

\begin{equation}
V(a)=V_0\; a^{-\lambda^2}+W_0\; a^{-(\alpha\lambda+3\gamma)},
\label{double}
\end{equation}
consequently, at early time (small $a$) the second term in the
right hand side (RHS) of (\ref{double}) prevails so, the first
term can be neglected and we can talk about a single exponential
potential with the constant $-(\alpha+3\gamma/\lambda)$ in the
exponent (see Eqn.(\ref{potencialdoble})). At late time ($a\gg
1$), instead, the first term in the RHS of (\ref{double})
dominates, so we can talk about a single exponential with the
constant $-\lambda$ in the exponent. For this reason, and owing to
the fact that in both cases (case A and case B), the coupling
function is of an exponential form, we focuss (as first
approximation) on the stability analysis of the model with a
single exponential potential of the form:
\begin{equation}
V=V_{0}\exp(-\bar a \phi),
\end{equation}
and the following coupling function:
\begin{equation}
X=X_{0}\exp(\bar b \phi).
\end{equation}
The constants $\bar a$ and $\bar b$ depend on the model. Using the
variables\cite{wands}
\begin{equation}
x\equiv\frac{\dot{\phi}}{\sqrt{6}H},\;
y\equiv\frac{\sqrt{V}}{\sqrt{3}H}
\end{equation}
the system of dynamical equations (\ref{frieman1}-\ref{frieman3})
can be arranged in the form of an autonomous dynamical system:

\begin{eqnarray}
x'=\bar a \sqrt{\frac{3}{2}}y^{2}-\bar b\sqrt{\frac{3}{2}}(1-
x^{2}-y^{2})\nonumber\\
-3x+\frac{3}{2}x[2x^{2}+\gamma(1- x^{2}-y^{2})] \label{estax}
\end{eqnarray}

\begin{equation}
y'=\frac{3}{2}y[2 x^{2}+\gamma(1- x^{2}-y^{2})]-\bar a
\sqrt{\frac{3}{2}}yx \label{estay}
\end{equation}
where, as before, the prime denotes derivative in respect to the
variable $N\equiv \ln(a)$. If in equations (\ref{estax}) and
(\ref{estay}) we set $\bar b=0$, the case studied in
Ref.\cite{wands} is recovered.

Using the new variables, the Friedmann equation looks like:

\begin{equation}
\frac{\rho_{m}}{3H^{2}}=(1-x^{2}-y^{2}),
\end{equation}
otherwise:
\begin{equation}
\Omega_{\phi}=x^{2}+y^{2}.
\end{equation}

Since $\rho_m\geq 0,$ the new variables are confined to the
compact region (see \cite{wands}), $0\leq x^{2}+y^{2}\leq 1$ so,
the evolution of the system under investigation is completely
described by trajectories within the unit disc. The lower
half-disc, $y<0$, corresponds to contracting universes, but we are
interest in expanding universes. On the other hand, due to the
following discrete symmetries : $(x,y)\rightarrow (x,-y)$ and
$t\rightarrow -t$, we consider the upper half-disc, $y\geq 0$ in
the subsequent discussion\cite{wands}.

The effective EOS parameter for the quintessence field at any
point is given by

\begin{equation}
\gamma_{\phi} \equiv
\frac{\rho_{\phi}+p_{\phi}}{\rho_{\phi}}=\frac{
\dot{\phi}^{2}}{V(\phi)+\dot{\phi}^{2}/2}=\frac{2 x^{2}}{
x^{2}+y^{2}}
\end{equation}

We are able to find up to five fixed points (critical points)
where $x'=0$ and $y'=0$, which are listed in Table \ref{crit}.

\bigskip

\begin{table*}[t]
\begin{center}
\begin{tabular}{|c|c|c|c|c|c|}
$x$ & $y$ & Existence & Stability & $\Omega_\phi$
 & $\gamma_\phi$ \\
\hline \hline -1 & 0 & All $\bar a$ and $\bar b$ & Stable node for
$\bar a <
-\sqrt{6}$ and &   1 & 2 \\
& & & $\bar b<\sqrt{\frac{3}{2}}(\gamma-2)$ & & \\
& & & Unstable node for $\bar a < -\sqrt{6}$ and & &\\
& & & $\bar b<\sqrt{\frac{3}{2}}(\gamma-2)$& & \\
& & & Saddle point otherwise& & \\
\hline 1 & 0 & All $\bar a$ and $\bar b$ & Stable node for $\bar a
> \sqrt{6}$ and & 1 & 2 \\
& & & $\bar b>\sqrt{\frac{3}{2}}(2-\gamma)$ & & \\
 & & & Unstable node for $\bar a <
\sqrt{6}$ and & & \\
& & & $\bar b<\sqrt{\frac{3}{2}}(2-\gamma)$ & & \\
& & & Saddle point otherwise & & \\
\hline $-\frac{\sqrt{\frac{2}{3}}\bar b}{\gamma-2}$ & 0 & All
$\bar a$ and $\bar b<\sqrt{\frac{3 (\gamma-2)}{2}}$ & Stable node
for $\bar a <\frac{3\gamma(2-\gamma)}{2\bar b}$ & $\frac{2\bar
b^{2}}{3(\gamma-2)^{2}}$ & 2\\& & & Saddle point for $\bar a
>\frac{3\gamma(2-\gamma)}{2\bar b}$ & &
\\\hline
$\frac{\bar a}{\sqrt{6}}$ & $\sqrt{1-\frac{\bar a^{2}}{6}}$ & All
$\bar a$ and $\bar b$ & Stable node for $\bar a^{2} < 6$ and $\bar
b>\frac{\bar a^{2}-3\gamma}{\bar a}$& 1 & $\frac{\bar
a^{2}}{3}$\\& & & Unstable node for $\bar a^{2} > 6$ and $\bar
b<\frac{\bar a^{2}-3\gamma}{\bar a}$ & &\\& & & Saddle point
otherwise& &
\\\hline
$\sqrt{\frac{3}{2}}\frac{\gamma}{\bar a-\bar b}$ &
$\sqrt{1-\frac{\bar a}{\bar a-\bar b}+\frac{3\gamma}{(\bar a-\bar
b)^{2}}-\frac{3\gamma^{2}}{2(\bar a-\bar b)^{2}}}$ & All $\bar a$
and $\bar b\leq \frac{\bar a^{2}-3\gamma}{\bar a}$ & Stable node
or Saddle point& $\frac{3\gamma+\bar b^{2}-\bar a\bar b}{(\bar
a-\bar b)^{2}}$ & $\frac{3\gamma^{2}}{3\gamma+\bar b^{2}-\bar
a\bar b}$
\\& & & (see analysis in table (\ref{crit2})) & &
\end{tabular}
\end{center}
\caption[crit]{\label{crit} Existence and stability of the
critical points.}
\end{table*}

\begin{table*}[t]
\begin{center}
\begin{tabular}{|c|c|}
Case 1 &  $0<\bar a<{\sqrt{6}}\wedge 0<\bar b<\bar a\wedge
0<\gamma <1-\gamma_1$
\\\hline
Case 2 & $\bar a={\sqrt{6}}\wedge\{0<\bar b<b_1\wedge 0<\gamma
<1-\gamma_1\vee b_1\leq \bar b<\bar a\wedge
 0<\gamma <\gamma_2\}$\\\hline
Case 3&
 ${\sqrt{6}}<\bar a<2 {\sqrt{2}}\wedge \{0<\bar b\leq b_0\wedge (0<\gamma
<1-\gamma_1\vee
 1+\gamma_1<\gamma <2)\vee
b_0<\bar b<b_1\wedge \nonumber (0<\gamma <1-\gamma_1\vee
 1+\gamma_1<
\gamma <\gamma_2)$\\
& $\vee
 \bar b=b_1\wedge
(0<\gamma <1-\gamma_1\vee 1-\gamma_1< \nonumber  \gamma
<\gamma_2)\vee
 b_1<\bar b<\bar a\wedge
 0<\gamma <\gamma_2\}$\\\hline
 Case 4 &
 $\bar a=2{\sqrt{2}}\wedge \{0<\bar b<b_0\wedge (0<\gamma <1-\gamma_1\vee
 1+\gamma_1<\gamma <2)\vee\nonumber$\\
 & $
 \bar b=b_0\wedge (0<\gamma <1-\gamma_1\vee
1-\gamma_1<\gamma <2)\vee b_0<\bar b<\bar a\wedge
 0<\gamma <\gamma_2\}$\\ \hline
Case 5 & $\bar a>2 {\sqrt{2}}\wedge\{0<\bar b<b_1\wedge
 (0<\gamma <1-\gamma_1\vee
 1+\gamma_1<\gamma <2)\vee
 \bar b=b_1\wedge
 (0<\gamma <1-\gamma_1\vee\nonumber$\\
 & $
 1-\gamma_1<\gamma <2)\vee
 b_1<\bar b\leq b_0\wedge
 0<\gamma <2\vee b_0<\bar b<\bar a\wedge
 0<\gamma <\gamma_2\}$
\end{tabular}
\end{center}
\caption[crit]{\label{crit2} Cases in which the point 5 is a
saddle point (otherwise it is an stable point). We have used
$\gamma_1=\frac{{\sqrt{3-2 \bar a \bar b+2\ {\bar
b^2}}}}{{\sqrt{3}}},$ $\gamma_2=\frac{1}{3}({\bar a^2}-\bar a \bar
b),$ $b_1=\frac{\bar a}{2}-\frac{1}{2} {\sqrt{-6+{\bar a^2}}}$ and
$b_0=\frac{-6+{\bar a^2}}{\bar a}$. $\wedge$ and $\vee $ are the
logical operators "and", "or" respectively.}
\end{table*}

\subsection{Double exponential potential}

\begin{figure}[t]
\centering
\leavevmode\epsfysize=5cm \epsfbox{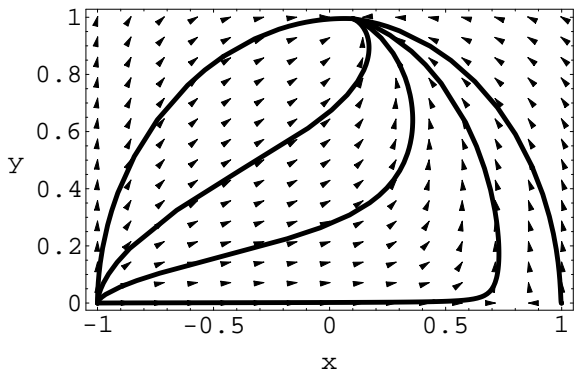}\\
\bigskip \caption[stabiA1]{\label{stabiA1} The phase plane for
$\gamma=1$, $\lambda=0.245$ and $\alpha=1$. The point ($0.1$,
$0.99$) is an attractor. All trajectories in phase plane diverge
from the unstable nodes at (1, 0) and (-1, 0).}
\end{figure}

\begin{figure}[t]
\centering
\leavevmode\epsfysize=5cm \epsfbox{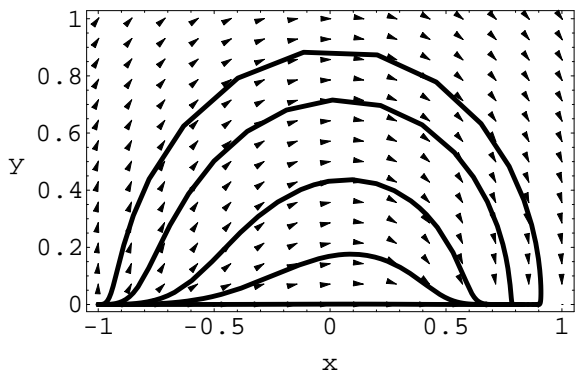}\\
\bigskip \caption[stabiA2]{\label{stabiA2} The phase plane for
$\gamma=1$, $\lambda=0.245$ and $\alpha=1$. The point ($0.81$,
$0$) is an attractor. All trajectories in phase plane diverge from
an unstable node (-1, 0).}
\end{figure}

In this case the constant $\bar b=\alpha$ is fixed. Since we have
used the constrain (\ref{restriccion1}), then the dynamical
variable $x$ is fixed to be $x=\frac{\lambda}{\sqrt{6}}$ so,
depending on $\bar a$, we are led with two possibilities: i) $\bar
a=\lambda$; the point 4 in table \ref{crit} is an attractor
quintessence dominated regime (Fig.\ref{stabiA1}) and ii) $\bar
a=\alpha+3\gamma/\lambda$; the point 5 in table \ref{crit} is a
matter-scaling attractor (Fig.\ref{stabiA2}). Therefore, we can
trace the evolution of the universe from a matter-scaling
attractor in the past into a late time, DE dominated attractor.

\subsection{Single exponential potential}

In this case two critical points emerge. Actually, although $\bar
a$ and $\bar b$ are fixed (we have a single exponential in this
case): $\bar a=\lambda$ and $\bar b=\lambda/2(2-\gamma)$, the
points 3 and 4 in table \ref{crit} are consistent with the
constrain (\ref{restriccion1}). The point 3 is a scaling solution
and the point 4 is a solution dominated by the scalar field.

As in the former case, the evolution of the universe transits from
a scaling attractor into a scalar field dominated attractor in the
future. This fact makes possible to deal with the fine tunning and
the coincidence problems\cite{luca,pavon}.

\section{Observational Testing}

Although we can not talk literally about observational testing of
the solutions found (we point out that the observational facts are
not yet conclusive in some cases and are not accurate enough in
others), in this section we shall study whether our solutions
agree with some observational constrains that are more or less
well established. The main observational facts we consider are the
following\cite{turner}:

1.- At present ($z=0$) the expansion is accelerated ($q(0) < 0$).

2.- The accelerated expansion is a relatively recent phenomenon.
Observations point to a decelerated phase of the cosmic evolution
at redshift $z=1.7$. There is agreement in that transition from
decelerated into accelerated expansion occurred at a $z \approx
0.5$ \cite{triess}.

3.- The equation of state for the scalar field at present
$\omega_\phi(\omega_\varphi)\sim -1$ (it behaves like a
cosmological constant). With a 95$\%$ confidence limit
$\omega<-0.6$\cite{ptw}.\footnotemark\footnotetext{Corasaniti and
Copeland \cite{cc} found that the determination of the third peak
in the BOOMERANG data limits the value today of equation of state
$-1 \leq \omega_{\varphi}(0) \leq -0.93$.}

4.- Although, at present, both the scalar (quintessence) field and
the ordinary matter have similar contributions in the energy
content of the universe ($\Omega_m(0)=1/3\;
\Rightarrow\;\Omega_\phi(0)=2/3$), in the past, the ordinary
matter dominated the cosmic evolution,
\footnotemark\footnotetext{A sufficiently long matter dominated
decelerated phase is needed for the observed structure to develop
from the density inhomogeneities \cite{sen}} meanwhile, in the
future, the quintessence field will dominate (it already
dominates) and will, consequently, determine the destiny of the
cosmic evolution.

5.- As stated before \cite{bhm}, nucleosynthesis predictions
claims that at $95\%$  confidence level. $\Omega_{\varphi} (1 MeV
\simeq z = 10^{10}) \leq 0.045$.

6.- During galaxy formation epoch \cite{mst}, around $z \approx
2-4$, the value of quintessence density parameter is
$\Omega_\varphi< 0.5$.

Now we proceed to "observationally" test the solutions found in
the cases studied in the former sections.

\subsection{Double exponential potential}

\begin{figure}[t]
\centering
\leavevmode\epsfysize=5cm \epsfbox{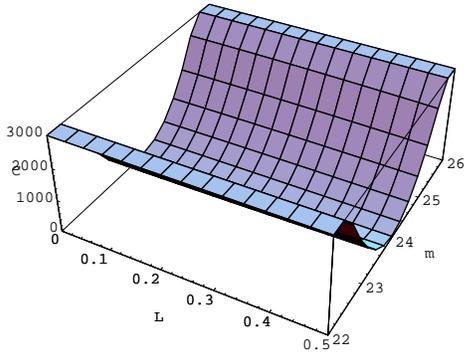}\\
\caption[chi]{\label{chi} The function $\chi^2$ ($C$ in the
figure) is plotted vs the free parameters $m$ and $\lambda$ ($L$
in the figure), for the model with double exponential potential.
We chose the parameter $\varepsilon=0.01$. Note that $\lambda$
could be any value in the physically meaningful range so, SNIa
luminosity observations do not allow for further constrain of the
parameter space. Other observations could be considered for this
purpose.}
\end{figure}

In this case the solution depends on two parameters, $\lambda$ and
$\varepsilon$ ($\alpha$ and $\gamma$ are fixed: $\gamma=1$,
meaning cold dark matter dominance at present, and $\alpha=1$). To
constrain the values of these parameters we use the aforementioned
facts (1-6). Using these constrains we bound the parameter space
($\lambda$ and $\varepsilon$) by means of a computing code. These
parameters range as $0\leq\varepsilon\leq 0.045$ and $0\leq
\lambda \leq 0.38$. Then we select a pair from the above interval
in order to compare our model with the supernova data. We follow
the procedure used in \cite{permu} to further constrain the
parameters in the model.\footnotemark\footnotetext{In fact this
procedure cannot be used to further constrain the space of
parameter, since the free parameters in our model are not very
sensitive to these observations.} For instance, if we consider
$\varepsilon=0.01$, the $\chi^{2}$ distribution has a minimum in
the $m_{0}$ direction\footnotemark\footnotetext{$m_{0}$ is a
parameter connected to the absolute magnitude and the Hubble
parameter.} at $m_{0}=24$; however, it has no minimum in the
$\lambda$-direction, meaning that the parameter $\lambda$ can take
any values in the interval $0\leq \lambda \leq 0.38$. In the
Figure \ref{chi}, $\chi^{2}$ is plotted as a function of the free
parameters $\lambda$ and $m_{0}$ (we chose $\varepsilon=0.01$,
$\lambda$ could be any value in the physically meaningful range).
Although the model fulfils the observational requirements (1-6)
given at the beginning of this section, we point out that other
observations should be considered in order to further constraint
the space of parameter.

\subsection{Single exponential potential}

In this case our solution depends on two parameters also $\lambda$
and $\epsilon$ (recall that we fix $\gamma$ to be unity). We apply
exactly the same procedure as that in the former subsection to
constraint the parameter space. We found that the physically
meaningful region in the parameter space is bounded by $0\leq
\epsilon \leq 0.045$ and $0\leq \lambda^{2} \leq 0.135$.

We want to stress that, as in the former case, the free parameters
of the model are not very sensitive to the SNIa data, so, the
standard used, for instance in \cite{permu}, is not suitable for
further constraining the space of parameters of the model. Other
observational evidence could be considered for this purpose.

\begin{figure}[t]
\centering
\leavevmode\epsfysize=5cm \epsfbox{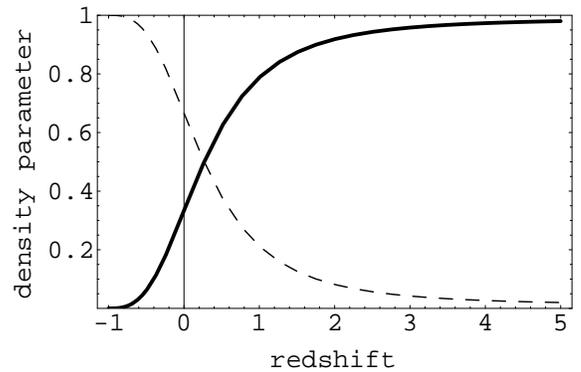}\\
\caption[densi]{\label{densi} The evolution of $\Omega_m$ (thick
solid line) and $\Omega_\phi$ (dashed line) vs $z$ is shown for
the model with a single exponential potential. The following
values of the free parameters $\varepsilon=0.01$ and $\lambda=0.3$
have been chosen. Equality of matter and quintessence energy
density occurs approximately at $z\approx 0.3-0.4$.}
\end{figure}

\begin{figure}[t]
\centering
\leavevmode\epsfysize=5cm \epsfbox{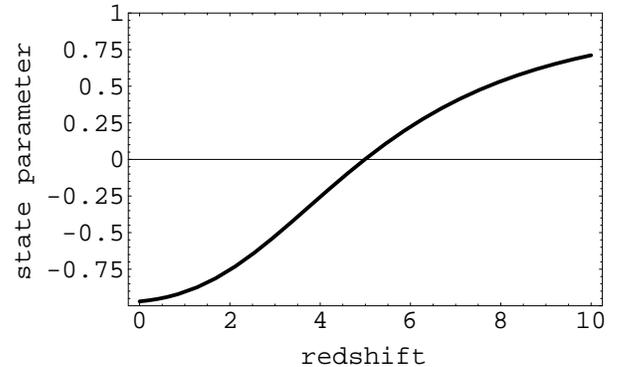}\\
\caption[state]{\label{state} We plot the dynamical EOS parameter
of the scalar field vs $z$ for the Model B (single exponential
potential). The values of the free parameters chosen are
$\varepsilon=0.01$ and $\lambda=0.3$ respectively. Note that the
DE evolves from almost being attractive matter (dust in
particular) in the past to behaving as an almost a cosmological
constant at present.}
\end{figure}

\begin{figure}[t]
\centering
\leavevmode\epsfysize=5cm \epsfbox{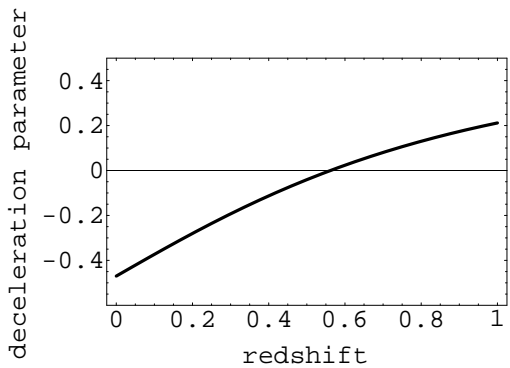}\\
\bigskip
\caption[des]{\label{des} We plot the evolution of the
deceleration parameter $q$ vs $z$ for the Model B (single
exponential potential). As before, the following values of the
free parameters $\varepsilon=0.01$ and $\lambda=0.3$ have been
chosen. The transition redshift when the expansion turns from
decelerated into accelerated occurred at $z_T\approx0.54-0.56$.}
\end{figure}

It should be pointed out that, in two cases (A and B), the
meaningful region in parameter space is chosen such that the main
observational facts (1-6) explained at the beginning of the
section are fulfilled. As an illustration, in Figure \ref{densi},
we show the evolution of both dimensionless DM and scalar-field
energy densities $\Omega_{m}$ and $\Omega_{\phi}$ respectively vs
$z$ for the single exponential potential (the Case A is very
similar). In Figure \ref{state}, the evolution of the equation of
state is shown vs $z$, while in Figure \ref{des}, we plot the
deceleration parameter to show the transition redshift when the
expansion turns from decelerated into accelerated
($z\approx0.55$).

\section{Conclusions}

We have found a new parametric class of exact cosmological scaling
solutions in a theory with general non-minimal coupling between
the components of the cosmic mixture: the cold dark matter and the
dark energy (the quintessence field). To specify the general form
of the coupling we were inspired in a scalar tensor theory of
gravity written in the Einstein frame. We have studied particular
coupling functions that lead to self-interaction potentials of the
following class: i) double exponential potential, ii) single
exponential potential. In order to derive exact flat FRW solutions
we have assumed a linear relationship between the Hubble expansion
parameter and the time derivative of the scalar field. The
stability and existence of these solutions have been studied.

In Case A (double exponential potential) the universe evolves from
an attractor regime, characterized by the scaling of matter and
dark energy, into a late time, dark energy dominated attractor
regime. In Case B we see that a single exponential can lead to one
of the following scaling attractor solutions\cite{barre}: either
1) a late-time attractor is a state when the scalar field mimics
the evolution of the background (DM) fluid, or 2) the late-time
attractor is the scalar field dominated solution. We show that the
assumed linear relationship between the Hubble parameter and the
time derivative of the scalar field always leads to attractor
scaling solutions.

In all cases, the models were tested against the widely accepted
observational facts. The space of parameters is then reduced to a
compact region whenever it is possible. It is shown that the
existing observational evidence coming from SNIa data is not
enough to further constrain the space of parameters. In all cases
the models fit with enough accuracy the observational data.

We conclude that models with non-minimal coupling between the dark
energy and the dark matter are easy to handle mathematically if
one assumes a suitable dynamics. Since, in the cases studied, the
coupling function is an exponential (these differ only in the
constant in the exponent), we point out that the resulting theory
belongs in the class of Brans-Dicke-type of theory, written in the
Einstein frame.

\section*{Acknowledgments}

We acknowledge the MES of Cuba by financial support of this
research.



\end{document}